\documentclass{article}
\usepackage{cite}
\usepackage{latexsym}
\usepackage{txfonts}
\usepackage{graphicx}
\usepackage{pxfonts}
\usepackage[british]{babel}
\usepackage{amsfonts}
\usepackage{authblk}
\newcommand{\be}{\begin{equation}}
\newcommand{\ee}{\end{equation}}
\newcommand{\gc}[1]{\left[#1\right]}
\newcommand{\gp}[1]{\left(#1\right)}
\newcommand{\Ps}{\Psi}
\newcommand{\pd}[2]{\frac{\partial #1}{\partial #2}}

\newcommand{\spd}[2]{\frac{\partial^2 #1}{\partial {#2}^2}}
\newcommand{\op}[1]{\frac{\partial}{\partial #1}}
\newcommand{\f}[2]{\frac{#1}{#2}}
\title{Ontological Motivation in Obtaining Certain Quantum Equations: A Case for Panexperientialism}
\author{Juan Villacr\'es \thanks{jvillacresb@gmail.com}}
\affil{Amazon Regional University Ikiam}
\date{}

\begin{document}
\maketitle

\begin{abstract}
In this work I argue for the existence of an ontological state in which no entity in it can be more basic than the others in such a state. This is used to provide conceptual justification for a method that is applied to obtain the Schr\"{o}dinger equation, the Klein-Gordon equation, and the Klein-Gordon equation for a particle in an electromagnetic field. Additionally, it is argued that the existence of such state is incompatible with indirect realism; and the discussion suggests that a panexeperientialist view is a straightforward means to embrace it. 
\end{abstract}

\section{Introduction}
\label{introduction}
This work begins with an analysis of the Madelung \cite{mad} equations, which are an alternative but equivalent formulation of the Schr\"{o}dinger equation.\footnote{Henceforth I refer to the Madelung equations as the `Schr\"{o}dinger equation' as has been done elsewhere, see e.g. \cite{reg}.} Specifically, I analyse a term of these equations that corresponds to the Fisher information, which suggests the existence of a state such that no entity in this state is more basic than the others in such a state -- this is done in section one. Although this result seems counterintuitive from an indirect realist perspective, the reasons to take seriously the existence of the mentioned state are ontological -- and are obtained in sections \ref{section3} and \ref{section4}.

The issues that are tackled in sections three and four can be seen as a means to shed some light on the following key questions. Is there a relation within the set of every entity that differs from another entity at a time $t_0$ that can hold between its elements and the ones corresponding to the set of every entity that is different from another entity at an immediate posterior time $t_1$?  Is it a metaphysical necessity that the existence of some entities should be more basic than others? In order to approach these questions, in section \ref{section3} I consider the sets of all entities (including relations) that differ from each other at an instant. I deploy an argument that shows that for any such set, there is no relation within it that can hold between its entities and those of another. Another similar argument is also developed and this leads us to the same conclusion, but this time with respect to atemporal sets of coexisting sets of the type just mentioned. In section \ref{section4} I then consider the possibility that the term `identity' refers to whatever makes an entity what it is, and which also rules out that such an entity should be other than what it is. The results of section \ref{section3} suggest that such a use of the term `identity' is granted by what reality can provide. This upshot reinforces our suspicions, raised in section \ref{section2}, about the existence of the mentioned state.

In spite of the apparently unrelated character of these questions, it is not surprising that these or similar questions should spring to mind when considering certain interpretations of the quantum gravity formalism, specifically those in which the status of spacetime as fundamental is disputed -- which in turn suggests its emergence \cite{unr,but,cal,lam,rom}. Notice that the possible lack of the mentioned status motivates the second question. On the other hand, when thinking about such emergence we can be led to think about the relation between the mentioned sets of different entities that include spacetime -- as is expressed in our first question.

Once we have strong reasons to assume the existence of the mentioned state, in section \ref{section5} I use its existence to justify a formal way to obtain: the Schr\"{o}dinger equation, the Klein-Gordon equation, and the Klein-Gordon equation for a particle in an electromagnetic field. In section \ref{section6}, I look for a view that can acknowledge the results of sections \ref{section3} and \ref{section4}. The discussion suggests that indirect realism is incompatible with such results, while on the other hand a panexperientialism similiar to one proposed by Bradley \cite{bra} seems suitable for embracing them. In section \ref{section7} I make some concluding remarks. 

\section{The Schr\"{o}dinger equation and Fisher Information}
\label{section2}
Let us begin by observing that the following expressions describe the classical (in contrast to quantum) behaviour of an ensemble of particles:\footnote{I work only in one dimension; however, the results can be generalized to more dimensions.}      

\be \pd{S}{t}+ \f{1}{2m}\gp{\pd{S}{x}}^2 + V=0 \label{hj}\ee

\be\pd{P}{t} + \f{1}{m}\op{x}\gp{P\pd{S}{x}}=0 \label{cont}\ee

where $S$ is action, $P$ is density of probability, $V$ is potential energy, $m$ is mass, $x$ is position, and $t$ is time. Equation (\ref{hj}) is the Hamilton-Jacobi equation,  while expression (\ref{cont}) is the probability conservation equation. These equations can be derived from the following functional by means of minimization.\footnote{That is, applying the Euler-Lagrange equation.}

\be L_1=\int P\gp{\pd{S}{t}+ \f{1}{2m}\gp{\pd{S}{x}}^2 + V} dxdt \label{L1}\ee

The central point of this section is that the Sch\"{o}rdinger equation can be obtained by applying minimization to

\be L_1 + \lambda \int\f{1}{P}\gp{\pd{P}{x}}^{2} dxdt \label{lq}\ee

with $\lambda=\frac{\hbar^{2}}{8m}$. In other words, applying minimization to (\ref{L1}) (as in the classical case), with the restriction that the following expression be fulfilled: 

\be \int\f{1}{P}\gp{\pd{P}{x}}^{2}dx=M \label{fi}\ee

where M is a constant. This being the case, this restriction seems to be the only difference between the mentioned classical equations and their corresponding quantum version. Now I analyse what this equation could be expressing. 

\subsection{Fisher Information}
The left part of Eq.(\ref{fi}),  
$\int\f{1}{P}\gp{\pd{P}{x}}^{2} dx$, is a measure of information. Specifically, it is a way of expressing the Fisher information, whose general form is: 

\[I=\int P(x|\theta)\gp{\pd{lnP(x|\theta)}{\theta}}^2dx\]

This gives a measure of information about the parameter $\theta$ that is present in the data $x$, where $P(x|\theta)$ is the density of the conditional probability of $x$ with $\theta$ given \cite{fri,cov}. 

We can assume that $x$ is the data obtained by measurement. For example, suppose that we are measuring a certain fixed position: the values that are obtained -- repeating measurements of the same position -- are the data $x$. We can consider $\theta$ as the value corresponding to the actual position.  Thus, in this case, the Fisher information is a measure of the amount of information that the data associated with our measurements carry about the data corresponding to the actual position. 

Let P be the proposition `it is not the case that every single measurement is equal to the \emph{actual} position', and let R be the proposition `there is at least one sample of measurements for which the Fisher information differs from the ones that are obtained for other samples'. We can assert that `if P, then R' is true. Equation \ref{fi}, however, expresses that no matter how many times we repeat the measurements, and irrespective of the values that are obtained, the amount of information that the data carry is constant. That is, Eq.(\ref{fi}) expresses that $R$ is false. That being so, we have it that $\neg$P is true. 

Essentially, the result of the analysis is that, no matter what values can be obtained by measurement, these also correspond to the actual position.  This implies that the category \emph{datum of a measurement of position} is the same as the category \emph{datum corresponding to an actual position}. This result seems to rule out that what an indirect realist refers to as a `mental representation' merely represents what actually happens.

Notice that an indirect realist might think that this result simply could not be right. She could argue that even in the case that a scientist intentionally uses a defective device to measure properties, the measurements obtained would also be the actual properties of the physical phenomena. That is, she could say you can never obtain a wrong measurement -- which is unacceptable.\footnote{And an immediate complaint that could arise from the indirect realist approach is that the cardinality of the extension of the expression `actual position of an object' differs from the one corresponding to the expression `data obtained from measuring the position of an object'.} Certainly, under the indirect realist view the ontological status of what corresponds to the \emph{actual} differs from the ontological status of what corresponds to a \emph{measure} -- the first being more basic than the latter -- which allows the possibility that a measure can be a wrong measure.\footnote{Notice that we cannot assert that an actual position is wrong. We can however make such an assertion when talking about a measure.} Therefore, I concede that the result is incompatible with indirect realism. This, however, does not mean that the result obtained in this section is wrong.  For, if it were the case that there is an ontological state in which any entity in it (including properties) has the same ontological status as the others, this result would just express that the actual position and its mental representations (using the indirect realist's idiolect) are in such a state -- as if such a way of being is instantiated by the property position and its mental representation. Maybe the existence of such an ontological state is what is being suggested by the Schr\"{o}dinger equation.  

At this point you may be wondering whether there are other ways to obtain the Schr\"{o}dinger equation by means of Fisher information, and whether these take an indirect realist approach. For, if that is the case, why should we bother considering the existence of the mentioned ontological state? Indeed, among the interpretations of the quantum formalism that are made by means of an analysis of information theory,\footnote{See e.g. \cite{whe,zei,fria,frib,shu}} there is Reginatto's `Derivation of the Equations of Non-relativistic Quantum Mechanics using the Principle of Minimum Fisher Information' \cite{reg} which uses the same mathematical expressions that I used above, and assumes indirect realism. Specifically, he considers that there is a mediator -- which introduces some noise -- between the actual property that is analysed and our measurements of it. Once he has made this stipulation, he assumes that our knowledge of the system that is going to be analysed is minimum. Hence, the epistemological aspect of his derivation `lies in the prescription to minimize the Fisher information associated with the probability distribution that describes the position of particles' \cite{reg}.  

In other words, Reginatto's derivation consists in applying minimization to Fisher information with the restriction that an ensemble of particles fulfills  equation \ref{cont}. That is, he is applying minimization to: 
\[\int\f{1}{P}\gp{\pd{P}{x}}^{2} dxdt + \lambda L_1\]
He suggests, however, that to do this is \emph{operationally equivalent} \cite{reg} to applying minimization to the restricted functional expressed in Eq.(\ref{lq}). That is, in the end he uses the formal method that I have already set out. The difference, it seems to me, is that Reginatto's approach is located within the framework of indirect realism, whilst, here, our conclusion that indirect realism is incompatible with the result obtained in this section is based on the analysis of Eq.(\ref{fi}) -- which also suggests the possibility of the existence of the mentioned state. 
I acknowledge, however, that such a suggestion is not strong enough for us take seriously the existence of such a state. Rather, and as we will see, its existence seems to be a consequence of the ontological issues treated in sections (3) and (4).

\section{Diversities that include space and time arise independently from other diversities}
\label{section3}
\begin{quote} `The question of relations is one of the most important that arise in philosophy, as most other issues turn on it \ldots idealism and realism, in some of their forms; perhaps the existence of philosophy as a subject distinct from science and possessing a method of its own' \cite{rus24} \end{quote} 

Let us consider relations of the type \emph{be different}. I will use $D_{ij}$ to refer to the relation that differentiates an entity $i$ from an entity $j$ (for any $i$ and $j$). Henceforth, the set of all entities (including relations) that differ from each other at an instant will be called an \emph{instant diversity}. My argument has the following premises: 

\begin{enumerate}

\item For any: entity $i$, $j$, relation $D_{ij}$, and instant; $i$ is different from $j$, and these are members of an \emph{instant diversity} if and only if $D_{ij}$ is a member of such a diversity.\footnote{Notice that the analysis presented in this section is committed neither to presentism nor non-presentism. For example, this \emph{instant diversity} could contain non-present existing objects, but this is not a requirement nor it is implicit in my argument. For this reason, the picture depicted here does not assume a preference for a Four-dimensionalist view (including the Stage view) nor for a Three-dimensionalist view. For an account of these views see e.g. \cite{sid}.} Note that every \emph{instant diversity} also comprises time, and the very instant associated with it.

\item For any instant, there is an \emph{instant diversity}\footnote{Note that in proposing this I am not requiring that relationism with respect to time be true. For premise $2$ does not exclude that an \emph{instant diversity} could contain e.g. just time, and an instant of time.} 

\end{enumerate}  

I use the term \emph{generation*} to refer to any relation that holds or that can hold between the entities (including relations) of an \emph{instant diversity} and the ones that belong to another \emph{instant diversity}. For example, suppose that a certain amount of water, certain blackberries, a certain blender, and an instant $t_1$ are members of the \emph{instant diversity} at that time. Now, imagine that we liquefy the mentioned water and blackberries in the blender. We would be tempted to say that the blackberry juice inside the blender -- which belongs to the \emph{instant diversity} at $t_2$ -- was \emph{generated*} by the mentioned amount of water, blackberries, and blender.           

I am going to prove by means of reductio ad absurdum that for any instant, there is no \emph{generation*} within the diversity corresponding to such an instant. Let $A=\{i, j, k, \ldots\}$ be the set of all entities (excluding relations) that are members of the \emph{instant diversity} at an instant $t_1$ -- where e.g. $i$ could be energy, $j$ could be space, $k$ could be time, etc. Also consider the set $R$ of all relations that are members of the mentioned \emph{instant diversity}. A subset of $R$ is the set of relations of the type \emph{be different} $B=\{D_{i\alpha}, D_{j\beta }, D_{k\gamma}, \ldots\}$; where $\alpha$ is the complement of the set $\{i\}$, $\beta$ is the complement of the set $\{j\}$ and so on. Notice that, due to premise $1$, $B$ is codependent with $A$.

Now, let us consider the first case where something, let us say $p$,\footnote{$p$ can be e.g. any entity, relation, an accidental property, an observable occurrence, etc.} is \emph{generated*} due to the relation $D_{p\delta}\in B$, where $\delta$ is the complement of $\{p\}$. Due to premise $1$, however, $D_{p\delta}$ is in $B$ if and only if $p$ was already a member of $A$ or $R$. Therefore, $D_{p\delta}$ is not \emph{generation*}. The other case is when there is already in $R$ a relation, let us say $G$, different from $D_{p\delta}$ that \emph{generates*} $p$. This implies that there is already in $B$ the relation of the type \emph{be different} $D_{GD_{p\delta}}$ between $G$ and $D_{p\delta}$. Now, applying premise $1$ to the relation $D_{GD_{p\delta}}$ and its relata, we have that $D{p\delta}$ was already a member of $B$. Therefore we return to the first case. That is, again, due to premise $1$, $D{p\delta}$ is in $B$ if and only if $p$ was already a member of $A$ or $R$. Therefore, there is no \emph{generation*} within the \emph{instant diversity} at $t_1$. This same analysis applies for each diversity corresponding to any instant. Therefore, for any instant, there is no \emph{generation*} within the diversity corresponding to such an instant.   

Note that this argument does not exclude that reality cannot provide, at least, a different \emph{instant diversity} from the one corresponding to this instant. For what was shown implies that we cannot find  \emph{within} any \emph{instant diversity} relations of the type \emph{be different} that can hold between it and another. In fact, there are two reasons why there can be, at least, a different \emph{instant diversity} from the one at this instant. Namely, in case it cannot exist, the entities characterized as \emph{the ones that I am aware of} -- e.g. my subjective mental experiences -- would be exhausted by the ones that I already have been aware of (which seem to be members of this \emph{instant diversity}). However, this is not the case. Furthermore, in the case of material objects, the impossibility of the existence of another \emph{instant diversity} would imply the ontological impossibility of obtaining new material objects apart from the ones that are members of the diversity at this instant.  This in turn excludes change in  ontological level for material objects -- which seems  not to be the case for our universe \cite{unr} \cite{kuc}.

Even if you defend a Parmenidean view -- e.g. based on an interpretation of the general covariance of general relativity as a gauge invariance \cite{unr} \cite{bel,kuc}  -- in which change is illusory, you would find problems. Illusory change seems to imply change only in what we perceive. The existence of such change implies that what is perceptible is not exhausted by what is already perceived. Because of this and due to what is already perceived seeming to be part of this \emph{instant diversity}, reality should be able to provide another \emph{instant diversity} aside from the one at this instant. 

Certainly, we can find relations between \emph{instant diversities}. As we have seen, reality can provide at least one \emph{instant diversity} that is \emph{different} from this \emph{instant diversity}. However, notice that causal relations, determination relations, or any relation you like between entities of different \emph{instant diversities} cannot be found within each of them. This being so, a way of thinking in which there is an underlying structure or law at $t_1$ that together with certain conditions -- initial ones -- at that time allows us to explain or to justify metaphysically the presence, absence or existence of some entity at a posterior time, seems to be misleading. Instead, it seems that such a law or structure can be provided by reality in case there is a broader set that can contain not just \emph{instant diversities} (or at least two of them), but also relations between their elements.

Also, at this point, you may perhaps be wondering about the way reality provides such different \emph{instant diversities}, considering that the elements of any of these cannot rely on the elements of another in order to obtain. In order to analyse this, firstly let us ask ourselves, do different \emph{instant diversities} coexist at an instant? For example, can the set of diverse entities (including relations) $C$ corresponding to $t_1$ coexist at such an instant with a different diversity $H$ that would correspond to $t_2$? It seems to me that the answer is no, because in that case the diversity at $t_1$ would be $C \cup H$, instead of $C$. Hence, it is not the case that reality provides a broader set of different \emph{instant diversities} at an instant. This being so, and acknowledging that no element of an \emph{instant diversity} can depend on others from another \emph{instant diversity} we are led to a picture in which this \emph{instant diversity} disappears and another comes into existence and comes to be present by replacing the former -- and simply when comparing these as if they were coexisting, relations between their elements can arise, as e.g. with the relation \emph{be different}. In such a case, however, relations that emerge between \emph{instant diversities} and between its elements could not be provided by observer-independent reality, but these seem to be required by our explanations when comparing or analysing more than one \emph{instant diversity}. Here, someone could complain that I am not considering an atemporal set of different entities which include every coexisting \emph{instant diversity}. In which case, change could be seen as actualizations of different existing elements of such set in a way that this \emph{instant diversity} ceases to be present and another replaces it.         

Let us consider such a set and call this an \emph{atemporal diversity}. Notice that an \emph{atemporal diversity} cannot contain two different \emph{instant diversities} as present entities. That is, the mentioned relation of actualization can hold just with one \emph{instant diversity} -- namely, with the present one. In order to see why I assert this, consider e.g. that the \emph{instant diversity} that contains the blackberries and water, and the one that contains the blackberry juice, exist. In case we want to assert that they are simultaneously present we concede that the blackberry juice is simultaneously present and not present. Here someone could complain that I am excluding the possibility that the blackberry juice be present in this world (or universe), and not present in some parallel world (or universe), as could be conceived in e.g. the Many Worlds interpretation of Quantum Mechanics \cite{eve}. This objection, however, does not affect my argument when applied just to this world (or universe). This being so, and due to the fact that what is present now does not exhaust what can be present, reality can provide at least one other \emph{atemporal diversity} which contains every \emph{instant diversity}, but in this case the one that is present differs from the one that was present in the former \emph{atemporal diversity}. 

Surely, you will now imagine that my next step is to deploy a similar argument to the one set out at the beginning of this section, in order to reveal what would be the relation, if any, between elements of different \emph{atemporal diversities}. Indeed, my premises are:

\begin{enumerate}

\item [1*] For any $i$, $j$, and $D_{ij}$; $i$ is different from $j$, and these are members of an \emph{atemporal diversity} if and only if $D_{ij}$ is a member of such diversity.

\item [2*] For any \emph{instant diversity} than can come to be present, there is an \emph{atemporal diversity}.

\end{enumerate}

When repeating the same argument as above -- but replacing \emph{instant diversity} with \emph{atemporal diversity}, the non-starred suppositions with the starred ones, and instead of considering different \emph{instant diversities} considering different \emph{atemporal diversities} -- we arrive at the conclusion that for any \emph{atemporal diversity} there is no \emph{generation**} within it. By \emph{generation**} I mean any relation that can hold or is held between elements of one \emph{atemporal diversity} and the ones of another. 

Notice that reality cannot provide a set of coexisting \emph{atemporal diversities} because this would imply that two different \emph{instant diversities} could be present. Because of this, and taking account that there is no relation within each \emph{atemporal diversity} that can link its elements with the ones of another, it seems that for \emph{atemporal diversities} it is necessary that when one diversity disappears another comes into existence, replacing the former independently of another \emph{atemporal diversity}. In this case, it seems that observer-independent reality is not providing relations between elements of different \emph{atemporal diversities} and between such diversities, but such relations seems to emerge as required by explanations when we analyse or consider more than one \emph{atemporal diversity}. This being so, when I talk about a \emph{different atemporal diversity} from the one that has this \emph{instant diversity} actualized, I refer to an \emph{atemporal diversity} that could emerge -- i.e. I am not referring to another coexisting \emph{atemporal diversity}. On the other hand, the mentioned process of disappearing and coming into existence, however, seems to not be necessary for \emph{instant diversities}. For in case reality can provide \emph{atemporal diversities}, it also can provide relations -- as elements of such \emph{atemporal diversities}, but not as elements of the \emph{instant diversities} -- that could link the elements of one \emph{instant diversity} with the elements of another.

\subsection{Objections to the idea that relations can be internal}

Objections could arise against premises $1$ and $1^*$. For it is implicit in these that $D_{ij}$ is closely linked with its relata in a way that the relation cannot be considered independent of its relata. In fact, we can see a strong dependency, since premises $1$ and $1^*$ imply that $i$ and $j$ -- as members either of an \emph{instant-diversity} or of an \emph{atemporal} one -- are a necessary condition for $D_{ij}$ to be a member of the mentioned diversities. In other words, in proposing the mentioned premises I am considering implicitly that relations of the type \emph{be different} are internal. Someone, however, could argue that Russell already showed that relations cannot be internal, in his `best-known, and probably most historically influential \ldots argument against Bradley' \cite{can}. 

Russell thought that Bradley defended the view that all relations are internal -- a view to which he refers as `monistic theory', and against which he demurs as we can see in the following passage:\footnote{However, as Candlish \cite{can} has pointed out, Bradley mainly defends the view that relations are unreal (see also \cite{spr}).} `The monistic theory holds that every relational proposition $aRb$ is to be resolved into a proposition concerning the whole which $a$ and $b$ compose \ldots The proposition ``a is greater than b'', we are told, does not really say anything about either $a$ or $b$, but about the two together. Denoting the whole which they compose by $(ab)$, it says we will suppose, ``$(ab)$ contains diversity of magnitude.'' Now to this statement \ldots there is a special objection \emph{in the case of asymmetry} [my emphasis]. $(ab)$ is symmetrical with regard to $a$ and $b$, and thus the property of the whole will be exactly the same in the case where $a$ is greater than $b$ as in the case where $b$ is greater than $a$' \cite{rus03}.

Notice that Russell's argument shows that the mentioned `monistic theory' cannot give a complete account of \emph{asymmetric} relational propositions. My argument, however, relies on the \emph{symmetric} relation of the type \emph{be different}. Hence, Russell's argument does not affect my analysis -- `$i$ differs from $j$' expresses the same as `$j$ differs from $i$'. Another opponent could argue that premise $1$ faces a problem when we apply Bradley's principle of fission \cite{bra} \cite{lem} to relations of the type \emph{be different} and to their relata. Let us see how such a principle works. 

\subsubsection{Bradley's principle of fission} 
We can assert on the basis of premise $1$ that a condition for $D_{ij}$ to be member of an \emph{instant diversity} is that $i$ be a member of such a diversity. To be in such a condition could be considered as a characteristic of $i$, which I will label as $a$. Also, from premise $1$ it seems that a consequence of $D_{ij}$ being a member of an \emph{instant diversity} is that $i$ be a member of it. Such a consequence could be considered as a characteristic of $i$, which I will label as $b$. Due to the fact that characteristics $a$ and $b$ are different -- which, due to premise $1$, implies that $D_{ab}$ is also a member of such a diversity -- they face the same analysis. This being so, $a$ would have two features of the same type -- i.e. one as a condition and the other as an effect -- and each of them another two characteristics of the same type, and so on, ad infinitum. In the words of Bradley `Each [relatum] has a double character, as both supporting and as being made by the relation' \cite{bra}. 

The opponent could advance her argument by asserting that this result imposes a restriction on the nature of the process of differentiation. Namely, when a differentiation is going to be performed for each step that is advanced, there is always a further new step that must be completed. Hence, there is no process of differentiation that could be fulfilled. In the words of Bradley, `We, in brief, are led by a principle of fission which conducts us to \emph{no end} [my emphasis]. Every quality in relation has, in consequence, a diversity within its own nature, and this diversity cannot \emph{immediately} [my emphasis] be asserted of the quality' \cite{bra}. From this, it seems that for something to be a member of an \emph{instant diversity} a constructive process of differentiation of characteristics has to be fulfilled -- where such a process has no end. 

I reply, however, based on premise $1$, that when a relation of the type \emph{be different} is a member of an \emph{instant diversity}, its relata \emph{simultaneously} are members of such a diversity. This being so, it does not matter whether infinite relata and relations of the type \emph{be different} should be required in order for something to be a member of an \emph{instant diversity}, because these are given together simultaneously -- in contrast to the purported constructive process -- or are not given at all. 
 
In the next section, I set out an analysis that together with the results obtained in this section supports our suspicion -- raised in the first section --  about the existence of a state such that no entity in it is more basic than the others in such state.

\section{The \emph{Symmetric State} and the \emph{Definite State}}
\label{section4}
\begin{quote}`If the explanation of ``there are Bs'' as meaning the same as ``Something that has being is B'' is to work, we just have to understand by being something that goes entirely without saying' \cite{fre}\end{quote}

Let us note that \emph{being} a certain entity implies \emph{not being another}. Formally, $\forall x \in \Gamma$ (if $x=x$, $\exists y\in \Gamma: x\neq y$), where $\Gamma$ is the set of all entities.\footnote{Here I am not going to discuss discernibility \cite{lad12}, I do not take into account either the Leibnizian Principle of Identity of Indiscernibles, or the Principle of Indiscernability of Identicals \cite{fre}. Additionally, I am not going to consider the `varieties of identity' as conceived by Martin-L\"{o}f \cite{lof} in his type theory \cite{ang}.} This suggests that we can conceive the identity of an entity as whatever makes it what it \emph{is} and which \emph{also} avoids it being another entity -- which can only be possible in case there are other such entities from which the former differs. Certainly, any such other will not be a certain unique entity -- instead, \emph{that} is whatever falls within the category \emph{other}. For an \emph{atemporal diversity} -- where, for each of its elements, its correspondent \emph{others} are given -- that an entity has its very identity (conceived as mentioned above) implies that -- as a metaphysical necessity -- it differs from every and each of the other members of such a diversity. Hence, when considering an entity X of a certain \emph{atemporal diversity} and taking its identity as mentioned before, that the other entities of such diversity be what they are seems to be a necessary condition for the identity (conceived as above) of $X$ being what it is. In case we conceive the identity of another entity $Y$ -- of the mentioned diversity -- as before, the identity of X would be a necessary condition for the identity of Y and vice versa. This being so, when the term `identity' as conceived above applies to more than one entity of the same \emph{atemporal diversity}, the relation of \emph{being a necessary condition} seems to be symmetric for them. In other words, an asymmetric relation of \emph{being a necessary condition} could not hold between them.  

Surely, this result seems incorrect, for an alteration or extinction of a certain entity -- e.g. the actual shoe that I am wearing -- of the \emph{atemporal diversity} that has this \emph{instant diversity} as present does not necessarily affect another entity -- e.g. the identity of the reader's shoe -- having the very identity that it has. Hence, it seems that my shoe does not have any ontological commitment that intersects with yours. This being so, an objector could assert that it seems that `identity' refers just to whatever makes an entity what it \emph{is} simpliciter, instead of referring to whatever makes an entity be the one that \emph{belongs} to a certain diversity -- which seems implied in my last analysis.  That is, it would seem to be a mistake to not consider identity isolated, independently from diversities. I concede that the objector's use of the term `identity' is guaranteed by what reality could provide; however, this does not invalidate the result obtained above. For the result of the last section implies that, for any diversity (either \emph{instant} or \emph{atemporal}), we cannot find within it the relation \emph{be the same} holding between its elements and the ones of another. It is necessary, however, that such a relation hold between entities of different diversities in order that whatever remains the \emph{same} across different diversities (or independently of them, if you want) could be the identity of an entity in the sense suggested by the objector. Due to this, the identity considered as whatever makes an entity be what it \emph{is}, which \emph{also} avoids it being another entity, seems to be guaranteed by what reality provides -- at least for the elements of each \emph{atemporal diversity}. For as was said before, there cannot be coexisting \emph{atemporal diversities}, and hence observer-independent reality cannot provide the mentioned relation the \emph{same} within a set that contains coexisting \emph{atemporal diversities}. Nevertheless, the identity considered as the one that remains the \emph{same} across (or independently of) \emph{instant diversities}, can also be supplied in case there are \emph{atemporal diversities}. For in this case, the relation the \emph{same} can be provided by observer-independent reality among the elements of an \emph{atemporal diversity} -- although not within each \emph{instant diversity}.

The conclusion of the analysis exposed hitherto in this section is that it seems that there are two uses of the term `identity' guaranteed by what reality can provide. It seems that the first use canvassed here guarantees  the existence of a special mode of being -- which I will call the \emph{symmetric mode}. For such a use, as was said before, when applied to an entity X of an \emph{atemporal diversity} implies that the other entities of such a diversity are necessary conditions for the identity of X. Furthermore, such a use suggests the existence of a special state\footnote{In this work the term `state' refers to states of affairs, but not in an `Armstrongian' sense.} -- which I will call the \emph{symmetric state} -- into which falls the group of entities (of the same \emph{atemporal diversity}) that present the \emph{symmetric mode} of being. For, as was said before, there is no asymmetric relation of \emph{being a necessary condition} that can hold between them, which suggests some ontological restrictions on entities that are in such a state. 

For example, it seems that if an entity $X$ is a fundamental brick of reality, an asymmetric relation of \emph{being a necessary condition} holds or can hold between this and at least some of the other entities. Suppose e.g. that matter and energy are fundamental bricks of reality as in some physicalist ontologies. This being so, there is an asymmetric relation -- of the mentioned type -- that holds or can hold between them and other entities, e.g. certain intentional mental phenomenon \cite{kim}. Our result would imply, however, that in case a group of entities that includes matter and energy are members of the same \emph{atemporal diversity} and these are in the \emph{symmetric state}, matter and energy would not be more basic than the other entities of the mentioned group. Hence, the \emph{symmetric state} is one in which all entities that are in it have the same ontological status.   
          
In order to get a better understanding of the character of the \emph{symmetric state} let us consider three of the four uses of the term `is' discussed by \cite{low09}, namely when `is' expresses instantiation, attribution and constitution. In general, it seems that the mentioned uses of `is' are applied to entities that have different ontological statuses -- as e.g. between kinds and objects or between a whole and its parts -- which is  impossible in the \emph{symmetric state}. That is, in general, it seems that such uses of `is' apply to entities that are not related by some symmetric relation of \emph{being a necessary condition}. For example, when asserting `This cow is white' there is at least a reading in which whiteness can exist even if the mentioned cow does not. Now, due to our result, propositions of the type `X \emph{is} S' -- with `is' used as was mentioned -- cannot be asserted  for entities in the \emph{symmetric state}, where the only characterizations allowed are the ones that assume symmetric relations \emph{of being a necessary condition} as e.g. `$a=a$'. 

Another interesting feature of the \emph{symmetric state} can be identified by considering that a property is intrinsic to an entity if the existence and nature of other entities `is counterfactually irrelevant' \cite{lad12} to the entity having the property. Notice that for entities in the \emph{symmetric state} there is no case in which the identity of each entity in the mentioned state is independent of the identity of each one of the other entities in such a state. Therefore, there is no entity in the \emph{symmetric state} that can be an intrinsic property of another entity in the mentioned state.           

Notice that in the  \emph{symmetric state} you can find whatever makes an entity what it \emph{is} (i.e. its identity); however, it seems that you cannot assert that `X is individuated by Y' for any pair $X$, $Y$ of different entities with identity that are in such a state. This is because in the mentioned state each entity helps another entity be what it is and not be any other. Nevertheless, `two different individuals cannot both individuate, or help to individuate, each other. This is because individuation in the metaphysical sense is a determination relation\ldots As such, individuation is an explanatory relation' \cite{low03} \cite{lad07}. Furthermore, there are no explanations (that is, none that assume asymmetric relations as being a necessary condition) that can be provided; instead, only explanations that assume the mentioned symmetric relations can be asserted, e.g. identity explanations. These explanations `cannot guarantee asymmetry' \cite{rub} as is required by causal explanations, but allow us to explain something conceptualized in one way by the same entity conceptualized in other way \cite{rub}.   

Certainly the other use of the term `identity' does not impose these ontological restrictions. For example, such a use is explicit in assertions such as: the specific egg and sperm from which I come are necessary for my identity \cite{kri}, but not vice versa. The lack of the mentioned restrictions when we consider that whatever remains the same across diversities could be the identity of an entity suggests that there is a wider state -- which I will call \emph{definite state} -- in which entities could be. In such a state entities can be fundamental, be characterizable by the mentioned asymmetric relations, and be explainable by such relations. 

I have given, I hope, good arguments to support the suspicion raised in the first section about the existence of a state such that there is no entity in it that can be more basic than the others -- namely, the \emph{symmetric state} -- or, at least, I have presented arguments in favour of its logical and metaphysical possibility. In the next section, I will show how the obtaining of some quantum equations is motivated and justified when we acknowledge that properties and their mental representations can be in the \emph{symmetric state}.

\section{About some Quantum equations}        
\label{section5}
\subsection{Justification of the Schr\"{o}dinger equation}

Recall that in section \ref{section2} we concluded that once we grant the existence of a state such that for any entity that is in it, such an entity is not more basic (ontologically speaking) than the others, then equation $\int\f{1}{P}\gp{\pd{P}{x}}^{2}dx=M$ (i.e. Eq.(5)) would express that the actual position (in the indirect realist idiolect) of the entity and -- what an indirect realist refers to as -- its mental representations are in such a state. Since the existence of such state seems to be granted by the results of the last section, the minimization of $L_1=\int P\gp{\pd{S}{t}+ \f{1}{2m}\gp{\pd{S}{x}}^2 + V} dxdt$ (i.e. expression (3)) with the restriction that Eq.(5) be fulfilled is required to express the case in which the mentioned position and its mental representations are in the \emph{symmetric state}. This procedure leads us to the Schr\"{o}dinger equation,\footnote{Alongside the mentioned process of minimization, we must also consider that the constant $\lambda=\frac{\hbar^{2}}{8m}$} which can be expressed as:      

\be \pd{S}{t} + \f{1}{2m}\gp{\pd{S}{x}}^2 -\frac{\hbar^2}{8m}\gp{\f{2}{P}\spd{P}{x}-\f{1}{P^2}\gp{\pd{P}{x}}^2} + V =0\label{sh1}\ee

\[\pd{P}{t} + \f{1}{m}\op{x}\gp{P\pd{S}{x}}=0\]

Notice that, if we consider that:

\[V_T=V -\frac{\hbar^{2}}{8m}\gp{\f{2}{P}\spd{P}{x}-\f{1}{P^2}\gp{\pd{P}{x}}^2}\]

equation (\ref{sh1}) can be expressed as the Hamilton-Jacobi equation:

\be \pd{S}{t} + \f{1}{2m}\gp{\pd{S}{x}}^2 + V_T=0 \label{jh2}\ee

However, the expression

\[V_Q=-\frac{\hbar^{2}}{8m}\gp{\f{2}{P}\spd{P}{x}-\f{1}{P^2}\gp{\pd{P}{x}}^2}\]

is not associated with any force or field, as is considered by interpretations such as the Hidden Variables interpretation \cite{boh,jam}, but rather appears as a consequence of acknowledging the existence of the \emph{symmetric state}.

\subsection{A method to obtain Quantum versions of Classical equations} 

The way in which the Schr\"{o}dinger equation was obtained in this proposal suggests that in order to obtain a quantum equation we should express formally that some property and its mental representation are in the \emph{symmetric state}, and require that this be the restriction of a functional of the form \[L_o=\int P\gp{H_o} dxdt\] when we minimize it. Where $H_o$ is the left side of a Hamilton-Jacobi equation,\footnote{Where such an equation is expressed in a way that zero is on the right hand side of the sign `='} we will see below that at least some of the quantum equations that could be obtained by this method are the quantum versions of the classical equation expressed by the mentioned Hamilton-Jacobi equation. Firstly, I am going to use this method for obtaining the Klein-Gordon equation. Notice that when doing this I will use formally the same procedure as Regginato (1998). The difference, however, lies in the conceptual framework that underlies each derivation. In our case, the acknowledge of the existence of the \emph{symmetric state} helps us to obtain, also, the Klein-Gordon equation for a particle in an electromagnetic field. 

\subsubsection{Application of the method to obtain the Klein-Gordon equation}
In the case that we take into account special relativity, the Hamilton-Jacobi equation has the following form: 

\[\gp{\pd{S}{x}}^2-\f{1}{c^2}\gp{\pd{S}{t}}^2+m_{o}^2c^2 = 0\]
Now, applying the method we have that in this case $L_o$ is:
\[L_{o}=\int P\gp{\gp{\pd{S}{x}}^2-\f{1}{c^2}\gp{\pd{S}{t}}^2+m_{o}^2c^2} dxdt\]
Now let us note that not only position but also time can share the \emph{symmetric state} with their respective mental representations. This being so, and in order to express that this is the case, the minimization of this functional $L_o$ must be restricted by requiring that the following equations be fulfilled:
\[\int \f{1}{P}\gp{\pd{P}{x}}^2 dx=M\]
\[\int \f{1}{c^2 P}\gp{\pd{P}{ t}}^2 dt=H\]
where M and H are constants. That is, we should apply minimization to:

\[L_{2}=\int P\gp{\gp{\pd{S}{x}}^2-\f{1}{c^2}\gp{\pd{S}{t}}^2+m_{o}^2c^2} + \f{\lambda}{P}\gp{\pd{P}{x}}^2 +\f{\beta}{c^2 P}
\gp{\pd{P}{ t}}^2 dxdt\]
From which we obtain:

\[\gp{\pd{S}{x}}^2-\f{1}{c^{2}}\gp{\pd{S}{t}}^2+ m_{o}^2c^2 - \gp{\f{2\lambda}{P}\spd{P}{x} + \f{2\beta}{c^2 P}\spd{P}{t}} + \gp{\f{\lambda}{P^2}\gp{\pd{P}{x}}^2 + \f{\beta}{c^2 P^2}\gp{\pd{P}{t}}^2}=0\]

\[ \op{x}\gp{P\pd{S}{x}}-\f{1}{c^2}\op{t}\gp{P\pd{S}{t}}=0\]
These equations with $\lambda = \frac{\hbar^2}{4}$, $\beta=-\frac{\hbar^2}{4}$ are equivalents to the Klein-Gordon equation,\footnote{The affirmation that such equations are equivalents to the Klein-Gordon equation can be verified if $\Psi=P^{1/2}exp(iS/\hbar)$ is replaced in the Klein-Gordon equation, and then the real and imaginary parts are separated. In this way we will obtain the mentioned equations. We can also reconstruct the Klein-Gordon equation starting from them.} which is usually expressed as: 

\[\hbar^2\spd{\Ps}{x}-m_{o}^2c^2\Ps=\f{\hbar^2}{c^2}\spd{\Ps}{t}\]

\subsubsection{Application of the method to obtain the Klein-Gordon equation for a particle in an electromagnetic field}
Now, following the proposed method we start from the Hamilton-Jacobi equation for a particle in an electromagnetic field, which is:
\[\gp{\pd{S}{x} -\f{e}{c}A_x}^2-\f{1}{c^2}\gp{\pd{S}{t}+e\phi}^2+m_{o}^2c^2=0\]
So, in this case:
\[H_o=\gp{\pd{S}{x} -\f{e}{c}A_x}^2-\f{1}{c^2}\gp{\pd{S}{t}+e\phi}^2+m_{o}^2c^2\]
Replacing this in $L_o$, and requiring that this be minimized taking into account -- as in the latter case -- that position, time and their mental representations are in the \emph{symmetric state}, the functional to minimize is:

\[L_3=\int P\gc{\gp{\pd{S}{x} -\f{e}{c}A_x}^2-\f{1}{c^2}\gp{\pd{S}{t}+e\phi}^2+m_{o}^2c^2} + \f{\lambda}{P}\gp{\pd{P}{x}}^2 + \f{\beta}{c^2P}\gp{\pd{P}{t}}^2 dxdt\]

Now applying minimization to $L_3$, from the variations with regard $P$ and $S$ we obtain:

\be \gp{\pd{S}{x} - \f{e}{c}A_x}^2-\f{1}{c^2}\gp{\pd{S}{t} + e\phi}^2 m_{o}^2c^2 + \gc{\f{\lambda}{P^2}\gp{\pd{P}{x}}^2+\f{\beta}{c^2P^2}\gp{\pd{P}{t}}^2}-\gc{\f{2\lambda}{P}\spd{P}{x}+\f{2\beta}{c^2P}\spd{P}{t}}=0\label{kge2}\ee

\be-\gc{\pd{P}{x}\gp{\pd{S}{x}-\f{e}{c}A_x} + P\gp{\spd{S}{x}-\f{e}{c}\pd{A_x}{x}}} + \f{1}{c^2}\gc{\pd{P}{t}\gp{\pd{S}{t} +e\phi} +P\gp{\spd{S}{t}+e\pd{\phi}{t}}}=0\label{kge3}\ee

The equations Eq.(\ref{kge2}) and Eq.(\ref{kge3}) with $\lambda=\f{\hbar^2}{4}$ $\beta=-\f{\hbar^2}{4}$ are equivalent to the Klein-Gordon equation for a particle in an electromagnetic field,\footnote{The affirmation that these equations are equivalent to the Klein-Gordon equation for a particle in an electromagnetic field is possible to verify if $\Psi=P^{1/2}exp(iS/\hbar)$ is replaced in it, and then the real and imaginary parts are separated. In this way, we obtain Eq.(\ref{kge2}) and Eq.(\ref{kge3})} which is usually expressed as:
\[\f{1}{c^2}\gp{i\hbar\op{t}-e\phi}^2\Ps=\gc{\gp{i\hbar\op{x}+\f{e}{c}A_x}^2 + m_{o}^2c^2}\Ps\]

Finally, note that a view that can support the existence of the \emph{symmetric state} can also motivate and justify the equations and the method analysed here. In the next section I briefly consider which view can support the results, and outline a view that requires that these results be fulfilled.

\section{A proposal}
\label{section6} 

The existence of the \emph{symmetric state} seems to be incompatible with indirect realism. For such view seems to require -- as a matter of metaphysical necessity -- the existence of entities that are more basic than others, as e.g. would be the case (in the mentioned view) between actual properties and their mental representations. On the other hand, should we embrace the results of this work while wanting to remain realists, it seems that we should look to direct realism for a view that can support our results. Indeed, as we will see, the panexperientialism proposed by Bradley \cite{bra} seems to have acknowledged a state close to the \emph{symmetric state}, which also seems to be not  incompatible with the results of section \ref{section3}. Due to this, I now outline a panexperientialist view by considering, firstly, that we have direct access to the constitutive material of reality. In my view, \emph{that} to which we have direct access is experience. Assuredly, you will recognize that I am referring to \emph{what} arises and could arise in the mind of any entity that has a mind: but please do not focus on the belonging relation between \emph{that} and the observer, but instead focus just on whatever arises or could arise in the mind. For when we take for granted that experience belongs to someone, and consider that it is the constitutive material of reality, we broach the problems associated with solipsism. 

I am going to assume that experience does not stand in need, in an ontological sense, of a mind to exist. Being so, by the term `experience' I am referring to observer-independent experience -- i.e. to \emph{what} arises and could arise in someone's mind, but which does not necessarily belong to someone.\footnote{Notice that, when accepting the existence of the \emph{symmetric state}, it is not the case that as matter of metaphysical necessity mental experiences belong to some observer. For there is the possibility that such an observer and the mental experiences be in the \emph{symmetric state}. This being so, in such a state we could not assert that it is true that `Some mental experiences belong to such an observer'. This is because the belonging relation seems to assume an asymmetric relation of \emph{being a necessary condition} between its relata.} Notice that, in a similar vein, Bradley had as a concern `to make it quite clear that this experience does not belong to any individual mind, and his doctrine not a form of solipsism' \cite{caba}.  In this sense, experience involves Fregean thoughts \cite{gas}, as e.g. the number $5$ which is something that can arise in your mind, but is not something that belongs to your mind. Experience also comprises perceptual experiences, \emph{qualia}, to which an indirect realist refers with the expression `mental representations', propositional attitudes, emotions, thoughts, abstract entities, and in general whatever is usually considered as mental experiences. It therefore seems to me that the following assumptions are fundamental for the proposal:

\begin{enumerate}

\item The reference and meaning of words are experiences. And it follows that the reference and meaning of the words `be' and `exist' are experiences.

\item The identity of any entity (including, for example, time, space, matter, energy, causality or the Self) is a set of experiences. 

For example, whether we identify an electromagnetic field through its properties or its dispositions \cite{bir}, these are certain abstract experiences taken from sensory data obtained through experiments. Its identity is the set whose elements are such experiences. 

\end{enumerate}

Note that when Bradley defends his panexperientialist view, he proposes  that the reader  extract all experience from anything that he/she considers has being and exists, and judges for him/herself whether it is still possible to speak about it or its being. He concludes: `I can myself conceive of nothing else than the experienced... Anything, in no sense felt or perceived, becomes to me quite \emph{unmeaning}' [my emphasis] \cite{bra}. Thus, it seems he thought that meaning is fundamentally related to experience, as I do in supposition $1$. Now, the results obtained in section \ref{section3} and the existence of the \emph{symmetric} and \emph{definite} state are also going to be conditions for this view, assuming that:

\begin{enumerate}
\item[3.] For any \emph{instant} or \emph{atemporal} diversity, there is no relation within it that can hold between its elements and those of a different diversity.   
\item[4.] The \emph{symmetric state} and the \emph{definite state} exist. 
\end{enumerate}
  
It seems to me that a view that fulfilled suppositions $1$ and $2$ is suitable for coherently embracing  conditions $3$ and $4$. In my opinion it is unproblematic, for example, to acknowledge in experience the existence of the \emph{definite state}. For instance, when `Socrates is wise' is true, the set of experiences that make up the entity referred to as `Socrates', and the set of experiences that make up the property referred to as `wise' are provided in a way that the latter is an attribute of the former -- where, as was said before, `is' used to express attribution seems to be applied to entities that are related by the asymmetric relation of \emph{being a necessary condition}. On the other hand, I also  consider that experience can be presented in a way that the mentioned sets of experiences named as `Socrates' and `wise' are not related by the mentioned asymmetric relations -- and that this applies to any set of experiences. In fact, Bradley's ontological view accepts the existence of states without any relation, which can be found `if you go back to mere unbroken feeling' \cite{bra}. It being the case that a panexperientialist view can support a state in which either symmetric or asymmetric relations are excluded, such a type of view seems to be able to acknowledge a state in which only the asymmetric relations of \emph{being a necessary condition} are excluded (i.e. it is able to support the \emph{symmetric state}).

Furthermore, a glimpse of Bradley's Absolute seems to reinforce our suspicion that a panexperientialist view would be suitable for supporting the results of this work. Let us consider such an Absolute as a `pre-conceptual state of immediate experience in which there are differences but no separations, a state from which our familiar, cognitive, adult human consciousness arises by imposing conceptual distinctions upon the differences' \cite{caba}. Allowing that this quotation does indeed summarize some important features of his Absolute, it seems that Bradley recognized two states. One, the pre-conceptual state of immediate experience, which seems to me close to the \emph{symmetric state}. In order to see this similarity, notice that here we consider that the constitutive material of reality is experience, and remember -- as was analysed in section \ref{section4} -- that in the \emph{symmetric state} entities do not have intrinsic properties, also they cannot be the relata of relations of attribution, constitution, instantiation, and individuation. Being so, it seems to me easy to be tempted to label such state as `pre-conceptual state of immediate experience'. The other state is such that experiences can be related in a way that allows reality to be characterized as we usually concede, which seems to me at least close to the \emph{definite state}.  For relations labelled as `separations' and `conceptualizations' extract some of the elements from the diversity known as the Absolute, and move them to another state in which experience is provided in a more familiar way.

Now, notice that in section \ref{section4}, condition $3$ was used to conclude the existence of the \emph{symmetric state}. That is, such a condition is sufficient for the existence of the mentioned state. Certainly, this is not enough to assert that a view that supports the existence of such a state also guarantees that condition $3$ is fulfilled. This, however, does not seem to contradict the other conditions. Furthermore, it seems that indirect realism would have trouble embracing such a condition. For the emergence of an \emph{atemporal diversity} independently of other entities -- which is a consequence of condition $3$ -- implies that there are no fundamental (ontologically speaking) entities involved in such emergence, and as was said before the existence of such fundamental entities seems a requirement for such view. Hence, the proposal seems to be suitable for supporting such a condition, or at least this view seems to be not incompatible with it.

\subsection{A couple of objections}

Someone could argue that, in order to deny something, I should be able to conceive of what I am denying. The objector could say that I am denying the existence of something that is not experience, and therefore I am experiencing something that is not experience. The dissenter could say that this contradiction implies that I cannot deny the existence of something that is not experience. To reply to this objection, first let us look at these thoughts more systematically. Let us label X the expression, `the existence of something that is not experience'. The premises of the objector's argument are therefore: 
\begin{enumerate}
\item If I deny X, then I conceive X
\item I deny X
\end{enumerate}

The conclusion reached through applying \emph{modus ponens} to these premises is: I conceive X. Considering that to conceive is to experience,  it follows that I am experiencing X. I completely agree that, in the case of the conclusion being false, one or all of the premises must be false; however, I neither affirm nor deny premise $2$, because X (interpreted, as the objector does, as referring to something that is not experience) is meaningless. That is to say, something is meaningful if it refers to experiences. In other words, the expression, `something that is not experience exists' is meaningful, and acquires some truth-value, if, and only if, such an expression refers to experiences. 

Another objector could claim that Putnam in his `Twin Earth' experiment \cite{put73,pes}, and Kripke in \emph{Naming and Necessity} \cite{kri} have shown that meaning and reference cannot be experiences. Instead, terms for natural kinds and names refer to the \emph{essence} of entities -- which in turn are not experiences. For example, he could say that `water' cannot fail to refer to its \emph{essence} -- which Putnam \cite{put73} took to be $H_2O$ -- and this is not experience. However, I reply, this objection seems to overlook what Putnam said about such a use of the word `essence': `But the \emph{essence} [my emphasis] of water in this sense is the product of our use of the word, the kind of referential intentions we have: this sort of \emph{essence} [my emphasis] is not `built into the world' in the way required by an essentialist theory of reference itself to get off the ground' \cite{put82}. Hence, it seems that our objector misunderstands the ontological essence of entities with the sort of essence about which Putnam actually talked in his `Twin Earth' experiment.

Furthermore, even accepting that terms for natural kinds,or names refer to ontological essences, the works of Putnam and Kripke mentioned by the objector do not prove that the constitutive material of such essences is not experience. For example, let us consider Putnam's motto ```meanings'' just ain't in the head' \cite{put73}. This refers to narrow mental content, e.g. subjective mental experiences, not to experience as I conceive it. That is, the Twin Earth thought experiment does not show that the constitutive material of essences, e.g. $H_2O$, is not experience. Additionally, in Kripke's work essence is what remains constant in all possible worlds, but such worlds are \emph{stipulated} counter-factual situations \cite{kri}. This being so, I can stipulate that what remains constant in all possible worlds is a set of experiences to which I refer as $H_2O$. Being so, terms for natural kinds and names can refer to experiences.

\section{Concluding Remarks}
\label{section7}

One of the main results of this work is that the \emph{symmetric state} exists. This is an ontological state for which any entity that falls into it is not more basic than the others in the mentioned state. This is a key element in the conceptual justification of the method used to derive the Schr\"{o}dinger equation. For when we use Fisher information to express the case in which the actual position and its mental representations are in the \emph{symmetric state}, we obtain the Schr\"{o}dinger equation from the classical equations condensed in the functional $L_1$. Furthermore, such formal method used to obtain the Klein-Gordon equation, and the Klein-Gordon equation for a particle in an electromagnetic field, is also justified and motivated by the acknowledge of the mentioned state. That is, the formal quantum expressions were motivated by an ontological commitment. So, a view that can acknowledge the mentioned state seems to be useful in Physics -- both for conceptual understanding and for formal developments. As was discussed before, indirect realism seems incompatible with the existence of such a state, whereas, on the other hand, an adequate candidate for supporting such a state is  the panexperientialist view proposed here  -- which was outlined in order to fulfil the results of this work.      

The other main result is that within each diversity (either \emph{atemporal} or \emph{instant}) there is no relation that holds, or that can hold, between the elements of such a diversity and those of another. This leads us to a view on which: in case there are no \emph{atemporal diversities}, where an \emph{instant diversity} disappears another comes to exist and to be present independently of that first \emph{instant diversity}; or in case that reality can provide \emph{atemporal diversities}, when the \emph{atemporal diversity} that has this \emph{instant diversity} actualized disappears, another comes to exist independently of that first \emph{atemporal diversity}. Also, the mentioned result seems to imply that in case \emph{atemporal diversities} exist, relations between \emph{instant diversities} and between their elements can be provided by observer-independent reality -- this in contrast with relations between \emph{atemporal diversities} and between their entities.

It seems that these results can be helpful not only for the conceptual and formal analysis of the quantum expressions explored here, but also to provide support in a natural way for the assertion that time is not fundamental -- as discussed on some interpretations of quantum gravity \cite{rom,lam,but,cal}. This is because accepting the existence of the \emph{symmetric state} implies acknowledging  one state in which time is not more basic than other entities in such a state. Note that the expression `emergence of time' used in such interpretations seems to refer to a relation that  holds between time and entities that are purportedly more basic than time --  although here such an expression would refer to the process of coming into existence of time, simpliciter. For, in this proposal, time comes into existence along with the other entities of each \emph{atemporal diversity} without depending on another entity. I gave, I hope, good arguments for considering the results obtained here and the framework that can support them -- in this case a panexperientialist one -- as being useful tools for work in Physics.

%\begin{acknowledgements}
%\end{acknowledgements}



\end{document}